\input mtexsis
\paper
\singlespaced
\widenspacing
\twelvepoint
\Eurostyletrue
\thicksize=0pt
\sectionminspace=0.1\vsize

\def\nc{{N}}
\def\yo1{{F_\pi^2}}

\def\oneht{\textstyle{1\over 2} }
\def\onehtsq{\textstyle{1\over{\sqrt{2}}} }

\def\oneft{\textstyle{1\over 4} }
\def\sss{\scriptscriptstyle}

\referencelist
\reference{gass}  J.~Gasser and H.~Leutwyler, 
\journal Ann. Phys.;158,142 (1984)
\endreference
\reference{poly}  M.V.~Polyakov, and V.V.~Vereshagin,
\journal Phys. Rev. D;54,1112 (1996)
\endreference
\reference{ecker}  G.~Ecker {\it et al},
\journal Nucl. Phys. B;321,321 (1989);
\journal Phys. Lett. B;223,425 (1989)
\endreference
\reference{*eckera}  For a review, see G.~Ecker, in {\it Chiral Dynamics: Theory and Experiment}, Proc. of the MIT workshop, 1994, eds. A.M.~Bernstein and B.R.~Holstein (Springer, 1995)
\endreference
\reference{sfsr}  S.~Weinberg, \journal Phys. Rev. Lett.;18,507 (1967)
\endreference
\reference{ope}  C.~Bernard, A.~Duncan, J.~LoSecco and S.~Weinberg, 
\journal Phys. Rev. D;12,792 (1975)
\endreference
\reference{beane1} S.R.~Beane, {{\tt hep-ph/9909571}}
\endreference
\reference{bijnens} For a review, see J.~Bijens, G.~Ecker and J.~Gasser, {\tt hep-ph/9411232}
\endreference
\reference{thooft} G.~'t Hooft,
\journal Nucl. Phys. B;72,461 (1974)
\endreference
\reference{*thoofta} E.~Witten,
\journal Nucl. Phys. B;156,269 (1979)
\endreference
\reference{bora} See, for instance, B.~Borasoy and Ulf-G.~Mei\ss ner, \journal Int.~J.~Mod.~Phys. A;11,5183 (1996), {\tt hep-ph/9511320}
\endreference
\reference{pdg}  Particle Data Group, \journal Eur.~Phys.~J.~C;3,1 (1998)
\endreference
\reference{derafael} E.~de Rafael and M.~Knecht, \journal Phys. Lett. B;424,335 (1998), {{\tt hep-ph/9712457}}
\endreference
\reference{pich} F.~Guerrero and A.~Pich,
\journal Phys. Lett. B;412,382 (1997), {{\tt hep-ph/9707347}}
\endreference
\reference{pipi} {{\it Pion-pion Interactions in Particle Physics}},
by B.R.~Martin, D.~Morgan and G.~Shaw, (Academic Press, London, 1976)
\endreference
\reference{alg}  S.~Weinberg, \journal Phys. Rev.;177,2604 (1969)
\endreference
\reference{buttiker}  B.~Ananthanarayan and P.~Buttiker, \journal Phys. Rev. D;54,1125 (1996), {\tt hep-ph/9601285}
\endreference
\reference{ulfm} B.~Borasoy and Ulf-G.~Mei\ss ner, \journal Ann. Phys.;254,192 (1997), {\tt hep-ph/9607432}
\endreference
\reference{mended}  S.~Weinberg, \journal Phys. Rev. Lett.;65,1177 (1990);
           {\it ibid}, 1181
\endreference
\reference{beane2} S.R.~Beane, \journal Ann. Phys.; 263,214 (1998), {{\tt hep-ph/9706246}};
\journal Phys. Rev. D;59,036001 (1999), {\tt hep-ph/9802283}
\endreference
\endreferencelist
\titlepage
\obeylines
\hskip4.8in{NT@UW-99-56}\unobeylines
\vskip0.5in
\title
Low Energy Constants from High Energy Theorems
\endtitle
\author
Silas R.~Beane

Department of Physics, University of Washington
Seattle, WA 98195-1560
\vskip0.1in
\center{{\it sbeane@phys.washington.edu}}\endcenter
\endauthor

\abstract
\singlespaced
\widenspacing

New constraints on resonance saturation in chiral perturbation theory
are investigated. These constraints arise because each consistent
saturation scheme must map to a representation of the full QCD chiral
symmetry group. The low-energy constants of chiral perturbation theory
are then related by a set of mixing angles. It is shown that vector
meson dominance is a consequence of the fact that nature has chosen
the lowest-dimensional nontrivial chiral representation. It is further
shown that chiral symmetry places an upper bound on the mass of the
lightest scalar in the hadron spectrum.

\endabstract 
\vskip0.5in
\center{PACS: 11.30.Rd; 12.38.Aw; 11.55.Jy; 11.30.Er} 
\endcenter
\endtitlepage
\vfill\eject                                     
\superrefsfalse
\singlespaced
\widenspacing

\vskip0.1in
\noindent {\twelvepoint{\bf 1.\quad Introduction}}
\vskip0.1in

The experimentally determined low-energy constants of chiral
perturbation theory are in excellent agreement with low-energy
constants determined by resonance saturation\ref{gass}.  This is no
surprise: the low-energy constants of chiral perturbation theory
represent the effect of resonances which have been integrated out of
the low-energy effective theory. What is surprising is that only a few
low-lying resonances account for all of the strength of the chiral
perturbation theory parameters. One might think it natural that the
lowest-lying states dominate. However, there is no separation of
scales in the spectrum which would indicate that only a given set of
low-lying resonances should dominate over all others at low energies.
Interestingly, large-$\nc$ arguments suggest that an infinite number
of resonances contribute with more or less equal strength, an
expectation which is realized in string-like models of $\pi -\pi$
scattering\ref{poly}.

Various theoretical constraints on resonance saturation have been
investigated\ref{ecker}. Foremost among these constraints are the
spectral function sum rules\ref{sfsr}, which are chiral symmetry
constraints on products of two QCD currents\ref{ope}. In a recent
paper it was shown that there are analogous sum rules for products of
three and four QCD currents\ref{beane1}. These sum rules also imply
constraints on resonance saturation. The purpose of this paper is to
consider all of the relevant constraints at once in the case of $\pi
-\pi$ scattering. There is then a simple chiral symmetry
interpretation of these constraints: all particles in a given saturation
scheme are in a representation of the full QCD chiral symmetry group
together with the pions\ref{beane1}. This interpretation clarifies resonance
saturation and sheds light on the ancient notion of vector meson
dominance.  

The reader may worry that because chiral symmetry is spontaneously
broken there is little sense in classifying states in the low-energy
theory using chiral symmetry. This worry is unfounded. A helpful way
to think is in terms of the operator product expansion, where it is
straightforward to prove that coefficient functions transform with
respect to the full global symmetry group of QCD, in spite of the fact
that this symmetry is spontaneously broken\ref{ope}. It is in fact
precisely this property of the operator product expansion which leads
to the sum rules which we study. The reader is referred to 
\Ref{beane1} for details.

In section 2 we write down the most general $SU(2)_L\times SU(2)_R$
invariant chiral lagrangian to order $p^4$ in the chiral limit. We
then consider the general theory of resonance saturation in section 3
and demonstrate the remarkable success of the vector meson dominance
picture. In section 4 we write down the complete set of chiral sum
rules relevant to $\pi-\pi$ scattering. We then consider the simplest
saturation schemes in section 5. We also derive an upper bound on the
mass of the lowest-lying scalar and consider explicit chiral symmetry
breaking effects. We conclude in section 6.

\vskip0.1in
\noindent {\twelvepoint{\bf 2.\quad Low-Energy Constants}}
\vskip0.1in

Consider the most general low-energy lagrangian consistent with the
symmetry breaking pattern $SU(2)_L\times SU(2)_R\rightarrow SU(2)_V$.
This low-energy lagrangian accommodates all underlying theories that
share this pattern of symmetry breaking. Well known technology tells
how to build the most general lagrangian involving pions consistent
with the relevant pattern of symmetry breaking\ref{bijnens}.  We
introduce a field $U$ that transforms linearly with respect to
$SU(2)_L\times SU(2)_R$: $U\rightarrow {L} U {R^\dagger}$ where
${{L,R}}$ is an element of $SU(2)_{L,R}$.  A convenient
parameterization of $U$ is

\offparens
$$
U=\exp{{i{\pi_a}{\tau_a}}\over{F_\pi}}
\EQN def1
$$\autoparens where $\tau_a$ are the Pauli matrices and $\pi_a$ is the
canonical pion field. The effective lagrangian describing the
interactions of the pion at low energies can be expressed as ${\cal
L}={\cal L}_2 + {\cal L}_4 +\ldots$ where the subscripts refer to the
number of derivatives.  The leading operator in the chiral limit is

\offparens
$$
{\cal L}_2=\oneft{F_\pi^2}{\it tr}\, ({D_\mu}U{D^\mu}U^\dagger ).
\EQN chirall2
$$\autoparens Here we use the covariant derivative
${D_\mu}={\partial_\mu}-i{r_\mu}U+iU{l_\mu}$ where ${r_\mu}$ and
${l_\mu}$ are external fields with associated non-abelian field
strengths $F_R^{\mu\nu}$ and $F_L^{\mu\nu}$, respectively.

At order $p^4$ there are four invariant operators in the chiral
limit\ref{gass}

\offparens
$$\eqalign{{\cal L}_4&= \oneft{{\it l}_{\sss 1}}
{\it tr}\, ({D_\mu}U{D^\mu}U^\dagger )^2
+\oneft{{\it l}_{\sss 2}}
{\it tr}\, ({D_\mu}U{D_\nu}U^\dagger ){\it tr}\, ({D^\mu}U{D^\nu}U^\dagger )\cr
&+{{\it l}_{\sss 5}}
{\it tr}\, ({F_{\mu\nu L}}U{F_R^{\mu\nu}}{U^\dagger})
-i{\oneht}{{\it l}_{\sss 6}}
{\it tr}\, ({F^{\mu\nu}_L}{D_\mu}U{D_\nu}{U^\dagger} +
{F^{\mu\nu}_R}{D_\mu}{U^\dagger}{D_\nu}{U} ) \cr}
\EQN chirall4
$$\autoparens where we have taken into account the coupling of the
pions to the external fields. The renormalization scale-dependent
parameters ${{\it l}_{\sss 1}}$, ${{\it l}_{\sss 2}}$, ${{\it l}_{\sss
5}}$ and ${{\it l}_{\sss 6}}$ are undetermined and independent insofar
as the pattern of symmetry breaking is concerned.  In the large-$\nc$
limit the continuum vanishes and the ${\it l}'s$ ---which are of order
$\nc$--- are determined by sums of infinite numbers of narrow
resonances\ref{thooft}.

\vskip0.1in
\noindent {\twelvepoint{\bf 3.\quad Resonance Saturation}}
\vskip0.1in

\vskip0.1in
\noindent {\twelvepoint{\it 3.1\quad Adding Matter}}
\vskip0.1in

It is straightforward to couple massive states with any quantum
numbers to the pions in a chirally invariant
way\ref{ecker}\ref{bijnens}. There are many ways of introducing
resonance fields. The basic rules of quantum field theory ensure that
all ways contribute the same physics at low energies\ref{bora}.

The axialvector couplings of the resonances to the pions are defined
by

\offparens
$$\eqalign{ 
&\bra{\pi_b}{Q^{5}_a}\ket{V_c}_i=
-i{\epsilon_{abc}}{G_{{\sss V_i}\pi}}/{F_\pi}\cr
&\bra{\pi_b}{Q^{5}_a}\ket{S}_i=
-i{\delta_{ab}}{G_{{\sss S_i}\pi}}/{F_\pi}\cr}
\EQN app1$$ 
where ${Q^{5}_a}$ are the QCD axial charges and $V$ and $S$ represent
states with $1^+(J^{--})$ ($J$ odd) and $0^+(J^{++})$ ($J$ even),
respectively (here we use the standard notation: $I^G(J^{PC})$). These
mesons are, respectively, the $\rho_{J}$'s with $J$ odd and the
$f_{J}$'s with $J$ even in the particle data tables.

The couplings to the vector and axialvector currents are given by:

\offparens
$$\eqalign{ 
&\bra{0}{A_{a\mu}}\ket{\pi_b}=
{\delta_{ab}}{F_{\pi}}{p_\mu}\cr
&\bra{0}{A_{a\mu}}\ket{A_{b}}_i^{\sss (\lambda )}=
{\delta_{ab}}{F_{\sss A_i}}{M_{\sss A_i}}{\epsilon_\mu^{\sss (\lambda )}}\cr
&\bra{0}{V_{a\mu}}\ket{V_{b}}_i^{\sss (\lambda )}=
{\delta_{ab}}{F_{\sss V_i}}{M_{\sss V_i}}{\epsilon_\mu^{\sss (\lambda )}}\cr}
\EQN app1$$ 
where and ${V_{a\mu}}$ and ${A_{a\mu}}$ are the (conserved) QCD vector
and axialvector currents and $\epsilon_\mu^{\sss (\lambda )}$ is the
vector meson polarization vector. Of course ${F_{\sss V_i}}\neq 0$ only
if $J=1$.  Here $A$ represents states with $1^-(1^{++})$. These mesons
are the $a_{\sss 1}$'s in the particle data tables.

\vskip0.1in
\noindent {\twelvepoint{\it 3.2\quad Vector Meson Dominance}}
\vskip0.1in

Consider saturating the low-energy constants with a single $V$ state, $\rho
(770)$, and a single $A$ state, ${a_{\sss 1}}(1260)$. We then have\ref{gass}\ref{ecker}

\offparens
$$\eqalign{ 
&{{\it l}_{\sss 1}}= -{{G_{\rho\pi}^2}\over{M_\rho^2}}\qquad 
{{\it l}_{\sss 2}}= {{G_{\rho\pi}^2}\over{M_\rho^2}}\cr
&{{\it l}_{\sss 5}}= -{{F_\rho^2}\over{4{M_\rho^2}}}
+{{F_{a_{\sss 1}}^2}\over{4{M_{a_{\sss 1}}^2}}}\qquad
{{\it l}_{\sss 6}}= -{{F_\rho}{G_{\rho\pi}}\over{{M_\rho^2}}}.\cr}
\EQN ressat1$$ 
Experiment gives 

\offparens
$$
{\yo1\over{G_{\rho\pi}^2}}
\simeq 1.9\ \qquad\quad
{{F_\rho^2}\over\yo1}
\simeq 2.7
\EQN data1$$ 
which are extracted from the decays ${\rho}\rightarrow{\pi}{\pi}$ and ${\rho^{\sss
0}}\rightarrow{e^+}{e^-}$\ref{pdg}.  The experimental situation can
then be roughly summarized through the relations:

\offparens
$${F_\pi}={\onehtsq}{F_{\rho}} \qquad
{G_{\rho\pi}}={\onehtsq}{F_\pi}.
\EQN oneegpd
$$ 
Equivalently we can determine $\sqrt{2}{G_{\rho\pi}}={F_\pi}$ from
${\rho}\rightarrow{\pi}{\pi}$ and then use the sum rule
${G_{\rho\pi}}{F_\rho} ={F_\pi^2}$\ref{ecker}.  The relations

$$
{F_\pi}={F_{a_{\sss 1}}} \qquad
{M_\rho}={\onehtsq}{M_{a_{\sss 1}}} 
\EQN twoegpd
$$ 
then follow directly from spectral function sum rules. The various
sum rules will be discussed in detail below.  With these values of the
resonance parameters the low-energy constants are related by:

\offparens
$$
-2{{\it l}_{\sss 1}}=2{{\it l}_{\sss 2}}
=-{8\over 3}{{\it l}_{\sss 5}}=-{{\it l}_{\sss 6}}
={{F_\pi^2}\over{M_\rho^2}}\equiv {\bar {\it l}}
\EQN fundequ
$$\autoparens 
and compared to experiment in Table 1. The agreement is rather
striking for such a simple saturation scheme. This is the modern
version of vector meson dominance (VMD).

\table{gl}
\tenpoint
\caption{The naive VMD scheme. Coefficients at one loop order in chiral perturbation theory taken
from \Ref{bijnens}, evaluated at $\mu ={M_\rho}$. The theoretical predictions are taken from 
\Eq{fundequ}}
\doublespaced
\ruledtable
$i$ | ${{\it l}_i^{exp}}\times 10^{-3}$ | Source | ${{\it l}_i^{th}}\times 10^{-3}$ \cr
$1$ | $-5.4\pm 2.5$  | $K_{e4};\, \pi\pi\rightarrow\pi\pi$ | $-7.3$   \cr
$2$ | $\;\; 5.4\pm 1.2$   | $K_{e4};\, \pi\pi\rightarrow\pi\pi$ | $\;\; 7.3$  \cr
$5$ | $-5.5\pm 0.7$  | $\pi\rightarrow e\nu\gamma$   | $-5.5$       \cr
$6$ | $-13.7\pm 1.4$ | $\vev{r^2}^\pi_{\sss V}$             |  $-14.6$  
\endruledtable
\endtable

\vskip0.1in
\noindent {\twelvepoint{\bf 4.\quad High-energy Theorems at Large-$\nc$}}
\vskip0.1in

The saturation scheme of section 3 demonstrates the phenomenological
success of the resonance saturation procedure when combined with
VMD. But is this scheme consistent with QCD? Generally, one may wonder
whether a saturation scenario with arbitrary particle content and
arbitrary masses and couplings is consistent with what we know about
QCD. This is {\it not} the case as there are important chiral symmetry
constraints which must be satisfied.

The chiral sum rules, or high-energy theorems, which must be satisfied
(in the chiral limit) are:

\offparens
$$\EQNalign{ 
&{\sum _{\sss V}}{F_{\sss V}^2}-{\sum _{\sss A}}{F_{\sss A}^2}=\yo1
\EQN summrelfir;a\cr
&{\sum_{\sss V}}{F_{\sss V}^2}{M_{\sss V}^{2}}-
{\sum_{\sss A}}{F_{\sss A}^2}{M_{\sss A}^{2}}=0
\EQN summrelfir;b\cr
&{\sum_{\sss V}}{F_{\sss V}}{G_{{\sss V}\pi}}=\yo1
\EQN summrelfir;c\cr
&{\sum_{\sss V}}{G^2_{{\sss V}\pi}}+{\sum_{\sss S}}{G^2_{{\sss S}\pi}}=\yo1
\EQN summrelfir;d\cr
&{\sum_{\sss V}}{G_{{\sss V}\pi}^2}{M_{\sss V}^2}-
{\sum_{\sss S}}{G_{{\sss S}\pi}^2}{M_{\sss S}^2}=0.
\EQN summrelfir;e\cr}
$$
\autoparens 
These sum rules contain the {\it totality} of the constraints which
the full chiral symmetry of QCD places on the low-energy constants of
chiral perturbation theory in the large-$\nc$ limit\ref{beane1}.  The
first two sum rules are familiar as the spectral function sum rules at
large-$\nc$\ref{derafael}. The other three sum rules are known to
follow from assumptions of unsubtracted dispersion relations for the
pion vector form factor\ref{ecker}\ref{pich} and the $I_t=1$ and
$I_t=2$ $\pi -\pi$ scattering amplitudes\ref{pipi}, respectively.  In
\Ref{beane1} it is shown that these five sum rules are exact in
large-$\nc$ QCD and follow directly from $SU(2)\times SU(2)$ symmetry.

\vskip0.1in
\noindent {\twelvepoint{\bf 5.\quad Finite Dimensional Saturation Schemes}}
\vskip0.1in

The chiral sum rules of \Eq{summrelfir} are saturated by an infinite
number of states in the large-$\nc$ limit.  In order to connect with
experiment we must consider saturation schemes with a finite number of
states. Chiral symmetry is respected provided that the sum rules are
satisfied.

\vskip0.1in
\noindent {\twelvepoint{\it 5.1\quad The Trivial Scheme}}

The simplest scheme contains $\pi$ and a single $V$ state, the $\rho
(770)$.  This corresponds to the six dimensional 
representation $(\bf{3},\bf{1})\oplus (\bf{1},\bf{3})$. The 
solution to the sum rules is

$${F_\pi}={F_\rho} \qquad {M_\rho}=0 \qquad
{G_{\rho\pi}}={F_\pi}. 
\EQN coneangstriv
$$\autoparens 
Of course, in this case, there is no mass splitting and chiral
perturbation theory can be consistent only if $\rho$ is kept as an
explicit degree of freedom, not a particularly interesting scenario.

Notice that the sigma model scenario ---containing $\pi$ and a single
$S$ state--- which corresponds to the four dimensional
$(\bf{2},\bf{2})$ representation, cannot satisfy all of the sum rules
and therefore is inconsistent with QCD, not a surprising result.  Of
course \Eq{summrelfir;d} is satisfied for {\it any} representation
involving $\pi$ and gives ${G_{{\sss S}\pi}}={F_\pi}$, which is, as 
expected, the (tree-level) sigma model value.

\vskip0.1in
\noindent {\twelvepoint{\it 5.2\quad The Simplest Nontrivial Scheme}}

The reader might have noticed that the VMD scheme of section 3 is not
consistent with the chiral sum rules. This is because that scheme
involves three isovectors, $\pi$, $\rho$ and $a_{\sss 1}$, which do
not fit into an $SU(2)\times SU(2)$ representation. Three isovectors
contain nine degrees of freedom. The only relevant representations of
$SU(2)\times SU(2)$ that contain isovectors are $\bf{(2,2)}$ and
$\bf{(1,3)\oplus (3,1)}$, which are dimension four and six,
respectively. There is simply no way to add four and six to give
nine. Therefore, consistency with chiral symmetry requires that the
saturation scheme include at least one additional isoscalar or one
additional isovector.  Hence it is clear that the simplest nontrivial
saturation scheme (i.e. with mass splittings) must contain $\pi$, $\rho
(770)$, $a_{\sss 1} (1260)$ {\it and} one $S$, which we take as
$f_{\sss 0}(400-1200)$\ref{alg}.  This corresponds to the ten
dimensional representation $(\bf{2},\bf{2})\oplus
(\bf{3},\bf{1})\oplus (\bf{1},\bf{3})$. Since the representation is
reducible there is a mixing angle. The solution of the sum rules is
given in terms of the mixing angle $\phi$:

$$\EQNalign{ 
&{F_\pi}={F_\rho}\sin\phi \qquad
{F_{a_{\sss 1}}}={F_\rho}\cos\phi \EQN coneangs;a\cr
&{M_\rho}={M_{a_{\sss 1}}}\cos\phi\qquad
{M_{f_{\sss 0}}}={M_\rho}\tan\phi\EQN coneangs;b\cr
&{G_{\rho\pi}}={F_\pi}\sin\phi \qquad
{G_{{f_{\sss 0}}\pi}}={F_\pi}\cos\phi .\EQN coneangs;c\cr}
$$\autoparens 
It is easy to verify that the sum rules are satisfied by \Eq{coneangs}: 

\offparens
$$\EQNalign{ 
&{F_{a_{\sss 1}}^{\sss 2}}+{F_\pi^{\sss 2}}=
{F_\rho^{\sss 2}} \EQN conesfsr;a\cr
&{M_\rho^{\sss 2}}{F_\rho^{\sss 2}}-
{M_{a_{\sss 1}}^{\sss 2}}{F_{a_{\sss 1}}^{\sss 2}}=0\EQN conesfsr;b\cr
&{G_{\rho\pi}}{F_\rho} ={F_\pi^2}\EQN conesfsr;c\cr
&{G_{\rho\pi}^2}+{G_{{f_{\sss 0}\pi}}^2}={F_\pi^2} \EQN conesfsr;d\cr
&{M_\rho^2}{G_{\rho\pi}^2}-{M_{f_{\sss 0}}^2}{G_{{f_{\sss 0}}\pi}^2}=0.
\EQN conesfsr;e\cr}
$$
Saturating the
chiral perturbation theory parameters with this particle content gives\ref{gass}\ref{ecker}

\offparens
$$\eqalign{ 
&{{\it l}_{\sss 1}}= -{{G_{\rho\pi}^2}\over{M_\rho^2}}
+{{G_{{f_{\sss 0}}\pi}^2}\over{2{M_{f_{\sss 0}}^2}}}\qquad 
{{\it l}_{\sss 2}}= {{G_{\rho\pi}^2}\over{M_\rho^2}}\cr
&{{\it l}_{\sss 5}}= -{{F_\rho^2}\over{4{M_\rho^2}}}
+{{F_{a_{\sss 1}}^2}\over{4{M_{a_{\sss 1}}^2}}}\qquad
{{\it l}_{\sss 6}}= -{{F_\rho}{G_{\rho\pi}}\over{{M_\rho^2}}}.\cr}
\EQN ressatme$$ 
The solution of the sum rules then yields

\offparens
$$\eqalign{ 
&{{\it l}_{\sss 1}}=-{\sss{1\over 2}}{\bar {\it l}}(\sin^2\phi -\csc^2\phi +2)
\qquad 
{{\it l}_{\sss 2}}={\bar {\it l}}\sin^2\phi \cr
&{{\it l}_{\sss 5}}=-{\sss{1\over 4}}{\bar {\it l}}(2-\sin^2\phi )
\qquad
{{\it l}_{\sss 6}}= -{\bar {\it l}}. \cr}
\EQN ressatmecasea$$ 

Now, we can fit the mixing angle to a datum. Say we determine
$\sqrt{2}{G_{\rho\pi}}={F_\pi}$ from ${\rho}\rightarrow{\pi}{\pi}$,
which implies $\phi\sim 45^0$.  We then have

\offparens
$$\EQNalign{ 
&{F_\pi}={\onehtsq}{F_{\rho}} \qquad
{G_{\rho\pi}}={\onehtsq}{F_\pi} \qquad
{G_{{f_{\sss 0}}\pi}}={\onehtsq}{F_\pi} 
\EQN egpdmod;a\cr
&{F_\pi}={F_{a_{\sss 1}}} \qquad
{M_\rho}={\onehtsq}{M_{a_{\sss 1}}}\qquad 
{M_\rho}={M_{f_{\sss 0}}}.
\EQN egpdmod;b\cr}
$$ Note that only a single datum has been used.  Chiral symmetry then
predicts all in terms of $F_\pi$ and $M_\rho$. In particular it is now
clear that the mysterious factors of $\sqrt{2}$ are related by chiral
symmetry. These relations are compared to experiment directly in Table
2.  This set of relations is to be compared to \Eq{oneegpd} and
\Eq{twoegpd}. 

Here the low-energy constants of chiral perturbation theory are related
through 

\offparens
$$
-4{{\it l}_{\sss 1}}=2{{\it l}_{\sss 2}}
=-{8\over 3}{{\it l}_{\sss 5}}=-{{\it l}_{\sss 6}}
={\bar {\it l}}
\EQN fundequ
$$\autoparens and are compared to experiment and to a Roy equation
analysis\ref{buttiker} in Table 3.  The sole difference with the VMD
scenario of section 3 is in ${{\it l}_{\sss 1}}$.  Evidently both
scenarios are compatible with data. We reiterate that in the method
advocated here, all resonances which contribute to the low-energy
constants are in a common chiral multiplet, while in \Ref{ecker} the
resonances are essentially decoupled from one another. It would be
interesting to reconsider the full three-flavor analysis of
\Ref{ecker} using $SU(3)\times SU(3)$.

\table{gl}
\tenpoint
\caption{Coupling and mass predictions taken from \Eq{coneangs} ---with (a) $\phi =45^0$ and
(b) $\phi =47^0$ (fit to ${G_{\rho\pi}}$)--- compared to experiment\ref{pdg}. We have also 
used $M_\rho =770$ and $F_\pi =93$. All numbers are in MeV. }
\doublespaced
\ruledtable
                    | TH (a,b)  | EXP  \dbl                          | TH (a,b) | EXP   \cr
${{F_\rho}}$ | $132,127$ | $153\pm 4$ \dbl ${{G_{{f_{\sss 0}}\pi}}}$ | $66,63$ |  $57-74$   \cr       
${{F_{a_{\sss 1}}}}$| $93,87$ | $122\pm 23$ \dbl ${{M_{a_{\sss 1}}}}$| $1089,1129$ |  $1230\pm 40$ \cr
${{G_{\rho\pi}}}$  | $66,{\it fit}$ | $68\pm 1$  \dbl ${{M_{f_{\sss 0}}}}$ | $770,826$  |  $400-1200$
\endruledtable
\endtable

\vskip0.1in
\noindent {\twelvepoint{\it 5.3\quad Is a Light Scalar Necessary?}}

Given the debatable status of the lowest-lying scalars it is
interesting to see whether it is possible to develop a realistic
scheme in which the lightest scalar mass is pushed up. We will see
that this is difficult to achieve unless some of the successes of the
VMD picture are sacrificed.

Assume that there is a single $V$ state, $\rho (770)$, and a single
$A$ state, ${a_{\sss 1}}(1260)$, and any number of $S$ states, and
further assume that $\phi =45^0$. The sum rules then
imply

\offparens
$$\EQNalign{ 
&{F_\pi}={F_{a_{\sss 1}}} \EQN egpdca;a\cr
&\sqrt{2}{M_\rho}={M_{a_{\sss 1}}} \EQN egpdca;b\cr
&{G_{{\rho}\pi}}={\onehtsq}{F_\pi} \EQN egpdca;c\cr
&{\sum_{\sss S}}{G^2_{{\sss S}\pi}}=\oneht\yo1 \EQN egpdca;d\cr
&{\sum_{\sss S}}{G_{{\sss S}\pi}^2}{M_{\sss S}^2}=\oneht\yo1{M_{\rho}^2}.
\EQN egpdca;e\cr}
$$ 
Multiplying \Eq{egpdca;d} by ${M_{\rho}^2}$ and subtracting
\Eq{egpdca;e} gives

\offparens
$$
{\sum_{\sss S}}{G^2_{{\sss S}\pi}}({M_{\sss S}^2}-{M_{\rho}^2})=0.
\EQN baproof
$$\autoparens which can hold only if there is at least one $S$ for
which ${M_{\sss S}}\leq{M_{\rho}}$. Hence chiral symmetry and VMD
place an upper bound on the mass of the lowest-lying scalar in the
hadron spectrum.  The inequality generalizes to ${M_{\sss
S}}\leq{M_{\rho}}\tan\phi$ for arbitrary $\phi$.

This argument relies on the assumption that the lowest-lying $S$ state
is in a chiral multiplet with the pion. A scalar in a separate
multiplet, for instance in the $(\bf{1},\bf{1})$ representation, would
be decoupled from the pion and hence would not contribute to the
low-energy constants of chiral perturbation theory.

\vskip0.1in
\noindent {\twelvepoint{\it 5.4\quad Explicit Breaking Effects}}

\table{gl}
\tenpoint
\caption{The consistent VMD scheme. Coefficients at one loop order in chiral perturbation theory 
taken from \Ref{bijnens}, evaluated at $\mu ={M_\rho}$. The theoretical predictions are taken 
from \Eq{fundequ}. Also included in the last column are Roy equation determinations of
${{\it l}_1}$ and ${{\it l}_2}$\ref{buttiker}.}
\doublespaced
\ruledtable
$i$ | ${{\it l}_i^{exp}}\times 10^{-3}$ | Source | ${{\it l}_i^{th}}\times 10^{-3}$ 
| ${{\it l}_i^{roy}}\times 10^{-3}$ \cr
$1$ | $-5.4\pm 2.5$  | $K_{e4};\, \pi\pi\rightarrow\pi\pi$ | $-3.7$ | $-5.4\pm 0.2$  \cr
$2$ | $\;\; 5.4\pm 1.2$   | $K_{e4};\, \pi\pi\rightarrow\pi\pi$ | $\;\; 7.3$ | $\sim 3.4$  \cr
$5$ | $-5.5\pm 0.7$  | $\pi\rightarrow e\nu\gamma$   | $-5.5$  | --     \cr
$6$ | $-13.7\pm 1.4$ | $\vev{r^2}^\pi_{\sss V}$             |  $-14.6$  | --
\endruledtable
\endtable

Taking into account a nonvanishing pion mass, \Eq{summrelfir;b} and 
\Eq{summrelfir;e} become:

\offparens
$$\EQNalign{ 
&{\sum_{\sss V}}{F_{\sss V}^2}{M_{\sss V}^{2}}-
{\sum_{\sss A}}{F_{\sss A}^2}{M_{\sss A}^{2}}={F_{\pi}^2}{M_{\pi}^2}
\EQN pisummrelfir;a\cr
&{\sum_{\sss V}}{G_{{\sss V}\pi}^2}({M_{\pi}^2}-{M_{\sss V}^2})-
{\sum_{\sss S}}{G_{{\sss S}\pi}^2}({M_{\pi}^2}-{M_{\sss S}^2})=0.
\EQN pisummrelfir;b\cr}
$$
\autoparens 
In the simplest nontrivial scenario this leads to the modified
mass relations,

$$\EQNalign{ 
&{M_\rho^2}={M^2_{a_{\sss 1}}}\cos^2\phi +{M^2_{\pi}}\sin^2\phi      
\EQN pimconeangs;a\cr
&({M^2_{f_{\sss 0}}}-{M_\pi^2})=({M^2_\rho}-{M_\pi^2})\tan^2\phi ,
\EQN pimconeangs;b\cr} $$\autoparens 
which imply (here we ignore the additional operators with explicit
breaking in chiral perturbation theory at order $p^4$) the
modifications:

\offparens
$$\EQNalign{ 
&{{\it l}_{\sss 1}}\rightarrow{{\it l}_{\sss 1}}-{\bar {\it l}}\;{{M_\pi^2}\over{2{M_\rho^2}}}
{\cot^4\phi}\;{\cos 2\phi} \EQN pimressatmecasea;a \cr
&{{\it l}_{\sss 5}}\rightarrow{{\it l}_{\sss 5}}+{\bar {\it l}}\;
{{M_\pi^2}\over{4{M_\rho^2}}}\cos^4\phi .
\EQN pimressatmecasea;b \cr}
$$ \autoparens Amusingly, the corrections to ${{\it l}_{\sss 1}}$
vanish for the choice $\phi =45^0$.  And the corrections to ${{\it
l}_{\sss 5}}$ are insignificant indeed:

$$
{{\it l}_{\sss 5}}=-{\bar {\it l}}\;
\Big\lbrace{3\over 8}-({{M_\pi}\over{4{M_\rho}}})^2\Big\rbrace .
\EQN mpifundequ
$$ These corrections are meant to be indicative of the size of
explicit breaking effects. As is usual in chiral perturbation theory
there are other effects arising at the next order ($p^6$) in the
chiral expansion which further shift the ${\it l}$'s~\ref{ulfm}.

\vskip0.1in
\noindent {\twelvepoint{\bf 6.\quad Conclusion}}
\vskip0.1in

The full $SU(2)\times SU(2)$ chiral symmetry of QCD places significant
constraints on resonance saturation in chiral perturbation theory.
Although some of these constraints have been studied previously, here
{\it all} of the constraints relevant to $\pi -\pi$ scattering have
been taken into account. We have found that the simple picture with a
single vector and a single axialvector state saturating the low-energy
constants of chiral perturbation theory is inconsistent with chiral
symmetry. This is easily seen by counting degrees of freedom and
matching to the dimensionality of allowed chiral representations.  In
particular, it would seem that chiral symmetry requires the presence
of isoscalar resonances. We have shown that the lightest scalar mass
is bounded above by $M_\rho$ if vector meson dominance is assumed.

According to the chiral symmetry point of view advocated here, vector
meson dominance is a consequence of the fact that in QCD, the pion
chiral representation is the lowest-dimensional nontrivial
representation, the ten dimensional
$\bf{(1,3)}\oplus\bf{(3,1)}\oplus\bf{(2,2)}$ representation, where the
angle $\phi$ which mixes the $\bf{(1,3)}\oplus\bf{(3,1)}$ and
$\bf{(2,2)}$ representations takes the value $45^0$.  Why the
$\bf{(1,3)}\oplus\bf{(3,1)}$ and $\bf{(2,2)}$ representations enter
with equal weight is mysterious and has been investigated in
\Ref{mended} and \Ref{beane2}.

\vskip0.15in


\noindent This work was supported by the U.S. Department of Energy grant
DE-FG02-93ER-40762. I thank Ulf Mei\ss ner, Dan Phillips and Mike
Strickland for valuable comments.

\vfill\eject 
\nosechead{References}
\ListReferences \vfill\supereject \end